\documentclass{iacrtrans}

\usepackage{float}
\usepackage[braket]{qcircuit}
\usepackage{tikz}
\usepackage{algorithm}  
\usepackage{algpseudocode}  
\usepackage{amsmath}  
\usepackage{cite}

\tikzset{every node/.append style={minimum size=0.75cm,draw,circle}}
\author{Anpeng Zhang, Xiutao Feng and Shengyuan Xu}
\institute{
 Academy of Mathmatics and Systems Science Chinese Academy of Science, Beijing, China, \email{{zhanganpeng, fengxt, xushengyuan18}@amss.ac.cn}
}

\title{Size optimization of CNOT circuits on NISQ}

\begin{document}

\maketitle

\keywords{CNOT circuits \and topologically-constrained \and graph structure \and MILP \and NISQ}

\begin{abstract}
  Quantum computers in practice today require strict memory constraints, where 2-qubit operations can only be performed between the qubits closest to each other in a graph structure. So a quantum circuit must undergo a transformation to the graph before it can be implemented. In this paper, we study the optimization of the CNOT circuits on some noisy intermediate-scale quantum(NISQ) devices. Compared with previous works, we decompose it into two sub-problems: optimization with a given initial qubit distribution and optimization without limitations of initial qubit distribution. We find that most of the previous researches focused on the first sub-problem, and ignored the influence of different distribution of qubits in the same topology structure on the optimization results. In this paper, We take both sub-problems into account and give some new optimization algorithms. In short, our method is divided into two steps: matrix optimization and routing optimization. We implement matrix optimization with the algorithm in \cite{liyi} and put forward a new heuristic algorithm with MILP method which can solve the second step. We implement our algorithm on IBM20 and some other NISQ devices, the results are better than most other methods in our experiment.

\end{abstract}

\section{Introduction}
With the rapid development of quantum computing technology, quantum computer has gradually become a reality. Now noisy intermediate-scale quantum (NISQ) \cite{preskill2018quantum} computers with 10-80 qubits have been made by some laboratories such as IBM and Google\cite{nay2019ibm}. 

In current quantum computers, many physical implementations of quantum computers such as superconducting quantum circuits \cite{nay2019ibm} \cite{reagor2018demonstration} and ion traps\cite{britton2012engineered} \cite{ballance2016high} impose strict memory limits, where 2-qubit operations (such as CNOT gates) can only be performed between the nearest qubits in a lattice or graphical structure. For a given quantum circuit with no limitation of memory operation, how to transform it into a circuit that can be realized on a real quantum computer has become an urgent problem to be solved. The traditional solution to this problem is to exchange qubits through the SWAP gate operation between adjacent qubits to achieve the realization of two quantum gates between two non-adjacent qubits\cite{zulehner2018efficient} \cite{herbert2018using}. However, this scheme is too costly, which usually increases the depth of the circuit and the number of gates by 1.5 to 3 times. In fact, finding the exact optimal solution appears intractable \cite{herr2017optimization}.

In 2019, Kissinger $et \ al$\cite{kissinger2019cnot} showed that a pure CNOT circuit can be compiled directly on an NISQ without significantly increasing the number of CNOT gates. Inspired by classic SLP problem, they transformed a CNOT circuit (linear map) into a matrix and then performed Restricted Gauss elimination to find its optimal implementation under the limit of some topological graphs. Later Nash $et \ al$ proposed a similar algorithm which can implement any CNOT circuit with a $4n^2$-size equivalent CNOT circuit on any connected graph\cite{nash2020quantum}. In 2021, Xiaoming Sun $et \ al$ put forward a new algorithm wihch gives a $2n^2$-size equivalent CNOT circuit for any CNOT circuit on any connected graph $G$ and proved that the lower bound of this problem is  $O(n^2/log \delta)$, where $\delta$ denotes the maximum degree of vertices in $G$\cite{2019arXiv191014478W}.

It has shown that finding the minimum CNOT gates in implementation for a CNOT circuit without restriction can  correspond to a Shortest Linear Program (SLP) problem, and the latter was proved to be an NP-hard problem \cite{boyar2013logic}. In 2020, Xiang Zejun $et \ al$ presented a new heuristic algorithm to search the optimal implementations of reasonable large matrices based on the s-Xor operation \cite{liyi}. Their basic idea is to find various matrix decompositions for a given matrix and optimize those decompositions to get the best implementation. In order to optimize matrix decompositions, they present several matrix multiplication rules over $F_2$, which are proved to be very powerful in reducing the implementation cost. 

In this paper we mainly discuss the optimization of the CNOT circuit under some topologically-constrained quantum memories. First we abstract it into a concrete mathematical problem, and give a new heuristic algorithm. Our algorithm includes two steps: 
\\
1) find the optimal implementation of the corresponding matrix of the CNOT circuit by Xiang $et \ al$'s method; 
\\
2) optimize the qubit distribution according to the result of Step 1) by means of the solver Gurobi.
\\
For consistency, we use the same set of randomly-generated CNOT circuits on 9, 16, and 20 qubits, and list the results in Table 1. The specific algorithm and implementation can be found on Github. It is seen that our results are better than the QuilC \cite{smith2016practical}, $t\ket{ket}$ compilers \cite{cowtan2019qubit} and stenier tree \cite{kissinger2019cnot} in most cases.

The paper is organized as follows. We start in Section 2 by reviewing some basic knowledge of quantum computing. In Section 3, we abstract the optimization of CNOT circuit into mathematical problems and detail our method, including matrix optimization and distribution optimization of qubits. In Section 4, we compared our experimental results with previous works and apply our method in the quantum implementation of AES’s Mixcolumns.
\section{Preliminaries}
\subsection{CNOT circuit}
Classical computer circuits consist of wires and logic gates \cite{nielsen2002quantum}. The wires are used to carry
information around the circuit, while the logic gates perform manipulations of the information, converting it from one form to another. For example, the logic gate Xor can transform bits $a$ and $b$ to $a\oplus b$, where $a,b=0 \ or \ 1$. But different from classical bit, the qubit is in superposition:
\[ \ket{\psi}=\alpha \ket{0}+\beta \ket{1} \]
And we can write it in a vector notation as 
$$ \ket{\psi}=\begin{bmatrix} \alpha \\ \beta \end{bmatrix}$$
In order to realize XOR operation on quantum computers, we introduce a multi-qubit gate, which is called the controlled-NOT or CNOT gate. This gate has two input qubits, known as the control qubit and the target qubit, respectively. 
\\
\begin{figure}[h]	
\centering
$$\Qcircuit @C=1.5em @R=1em {
&\ket{A} \qquad & \ctrl{1} & \qw  &\qquad \ket{A}\\
&\ket{B} \qquad & \targ & \qw  &\qquad \ket{A\oplus B}\\
}$$
\caption{Controlled-NOT gate}\label{fig:1}
\end{figure}
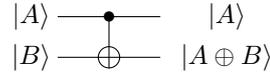
\\
Another way of describing the action of the CNOT gate is to give a matrix representation. In fact, every qubit gate can correspond to a unitary matrix, we use $U_{CN}$ to denote the CNOT gate. 
$$ U_{CN}= \begin{bmatrix} 1&0&0&0 \\ 0&1&0&0 \\ 0&0&0&1\\ 0&0&1&0 \end{bmatrix}$$
One can easily check that $U_{CN}^{\dag}U_{CN}=I$. Of course, there are many interesting quantum gates other than the controlled-NOT. But here we only focus on quantum circuit which is composed of CNOT gates.
\\
\\
\textbf{Definition 1}: A quantum circuit is called a CNOT circuit if it contains only CNOT gates.
\\
\begin{figure}[h]
\centering
\includegraphics[scale=0.6]{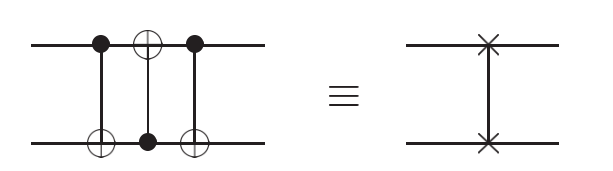}
\caption{An example of CNOT circuit swapping two qubits, and an equivalent schematic symbol notation} \label{fig:2}
\end{figure}
\\
In fact, since the CNOT circuit is linear, it is easy to map it to an $n\times n $ matrix over $GF (2)$, and each CNOT gate can correspond to an elementary row operation of a matrix, such as $CNOT_{i,j}$ is actually adding the row $i$ of the matrix to the row $j$.

\subsection{CNOT basis transformations}
In some practical quantum computers such as IBM-QX5 in figure 3, there is a strict limit between the control bit and the target bit, and the target bit cannot be used as the control bit for controlled operation, but if we only consider CNOT circuits, we can ignore the limitations by the follow fact.
\begin{figure}[h]
\centering
\includegraphics[scale=0.6]{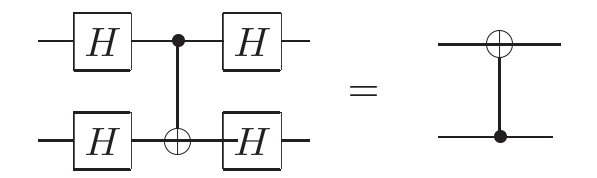}
\caption{a quantum circuit interchanging controlled bit and target bit of a CNOT gate} \label{fig:CNOT transformation}
\end{figure}
\\
Thus, with four hamdamard gates, we implement the interchange of a controlled bit and a target bit. All graphs we will mention in the rest of this paper are undirected graphs.
\begin{figure}[h]
\centering
\includegraphics[scale=0.6]{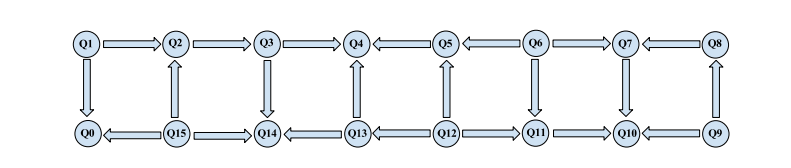}
\caption{The structure of IBM-QX5, where an arrow from qubit A to qubit B means that A can act as control qubit of a CNOT gate with target qubit B}
\end{figure}

\section{Optimization of the CNOT circuit}
Noisy intermediate-scale quantum (NISQ) has many different architectures, such as 3 × 3 and 4 × 4 square lattices, 16-qubit architectures of the IBM QX-5 and Rigetti Aspen devices, and the 20-qubit IBM Q20 Tokyo architecture. For each architecture, we can represent it as a graph $G(V,E)$, and CNOT operations are typically only possible between pairs of adjacent systems. In other words, we can only perform row operation between specific rows when doing Gauss-Jordan elimination. 
\\
\\
\textbf{Definition 2}: Given a graph $G(V,E)$ and the order $1 \sim |V|$ of vertices in $G$. $S(G)$ is a set of elementary matrices in which each element is an elementary row operation $E(i+j)$ where $E(i+j)$ is the resulting matrix by adding the $j$th row to the $i$th row of an identity matrix and labels of two rows are the ends of an edge in $G$.
\\
\\
Then the optimization of CNOT circuits can be expressed as the following questions.
\\
\\
\textbf{Problem 1}: Given a matrix $A$ over $GF(2)$ and $S(G)$ of a graph $G(V,E)$, how to decompose $A$ into the product of elements in $S(G)$ with the least elements.
\\
\\
One can easily check that different orders of $V$ of a same graph $G$ will derive different $S(G)$ and lead to different matrix optimization implementations. Naturally, we consider the following question.
\\
\\
\textbf{Problem 2}: Given matrix $A$ over $GF(2)$ and graph $G(V,E)$, find an order of $V$ and some matrixes in $S(G)$ such that we can decompose $A$ into the product of the least elements in $S(G)$.
\\
\\
Problem 2 is a generalization of Problem 1, and Steiner tree method in \cite{kissinger2019cnot} is for Problem 1, without considering the distribution of qubits. Unfortunately, since Problem 2 involves two critical factors including distribution of qubits and matrix optimization implementation, and these two factors are mutually restricted, the distribution of qubits will greatly affect the results of matrix optimization implementation, and the results of matrix optimization implementation will also affect the selection of qubit position. Therefore, it is difficult to optimize the two aspects at the same time. In this case, without considering the limitation of the topology structure, we firstly optimize the matrix, and then optimize the qubit distribution through the result of matrix optimization implementation.
\\
\subsection{Main methods}

\subsubsection{Matrix optimization implementation}
For each matrix $A$ over $GF(2)$, we perform Gauss-Jordan elimination on $A$, and each elementary row operation corresponds to some elementary matrix,
we will obtain $A_1A_2...A_kA= I$, where $A_i$ is
an elementary matrice. Reverse the order of $A_i$ we can represent $A$ as product of some elementary matrices.\\
Since the decomposition of a matrix is not unique, there is a natural problem that how to find the least elementary row operations to perform Gauss-Jordan elimination. This problem corresponds to the Shortest Linear Program (SLP) problem. As for this problem, we can follow some rules of reduction to find the optimal implementation\cite{liyi}. Here is an example.
\\
Let
$$ A=
\begin{bmatrix} 
1 & 1&0&1&1 \\ 0 & 0  & 1 & 1&0\\   1 & 0 & 1 & 0 & 1  \\ 1 & 1 & 0 & 1 & 0 \\ 1 & 1 & 1 & 1 & 0  
\end{bmatrix}
$$
\\
First we can represent A as product of some elementary matrices $E(i+j)$ and $E(i\leftrightarrow j)$, where $E(i+j)$ is the resulting matrix by adding the $j$th row to the $i$th row of an identity matrix, $E(i\leftrightarrow j)$ is the resulting matrix by exchanging the $i$th and $j$th row of an identity matrix.
\\
Then find the optimal implementation according to the following rules.
\begin{table}[h]
\centering
\begin{tabular}{|c|c|}
\hline 
\textbf{R1}& $E(k + i)E(k + j)E(i + j) = E(i + j)E(k + i)$ \\ \hline
\textbf{R2}& $E(i + k)E(k + j)E(i + j) = E(k + j)E(i + k)$ \\ \hline
\textbf{R3}& $E(i + k)E(j + k)E(i + j) = E(i + j)E(j + k)$ \\ \hline
\textbf{R4}& $E(j + k)E(i + k)E(i + j) = E(i + j)E(j + k)$ \\ \hline
\textbf{R5}& $E(k + j)E(k + i)E(i + j) = E(i + j)E(k + i)$ \\ \hline
\textbf{R6}& $E(k + j)E(i + k)E(i + j) = E(i + k)E(k + j)$ \\ \hline
\textbf{R7}& $E(j + i)E(i + j) = E(i\leftrightarrow j)E(j + i)$ \\ \hline
\end{tabular}
\end{table}
\\
We get
$$ A=
A'
\times
E(2+4)\times
E(4+3)\times
E(2+1)\times
E(1+5)\times
E(1+3)\times
E(5+2)\times
E(3+2)
$$
Where $A'$ is a permutation matrix
$$\begin{bmatrix} 
0 & 0&0&0&1 \\ 0 & 0  & 0 & 1&0\\   1 & 0 & 0 & 0 & 0  \\ 0 &1 & 0 & 0 & 0 \\ 0 & 0 & 1 & 0 & 0  
\end{bmatrix}$$
Then we get a CNOT circuit of A in Figure 5.
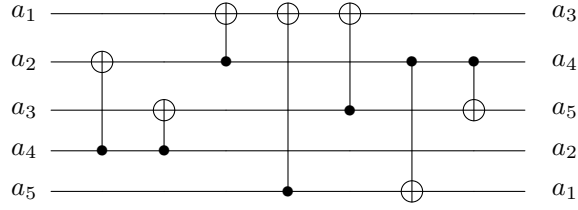
\begin{figure}[h] 
\centering
$$\Qcircuit @C=1.5em @R=1em {
&a_1 \qquad & \qw       & \qw       &\targ      & \targ     & \targ     & \qw      & \qw      & \qw &a_3 \\
&a_2 \qquad & \targ     & \qw       & \ctrl{-1} & \qw       & \qw       & \ctrl{3} & \ctrl{1} & \qw &a_4\\
&a_3 \qquad & \qw       & \targ     & \qw       & \qw       & \ctrl{-2} & \qw      & \targ    & \qw &a_5\\
&a_4 \qquad & \ctrl{-2} & \ctrl{-1} & \qw       & \qw       & \qw       & \qw      & \qw      & \qw &a_2\\
&a_5 \qquad & \qw       & \qw       & \qw       & \ctrl{-4} & \qw       & \targ    & \qw      & \qw &a_1\\
}$$

\caption{the CNOT circuit of $A$ without topological constraints}\label{fig:1}
\end{figure}
\subsubsection{Distribution optimization of qubits}
In order to describe the optimization of CNOT circuits as a quantifiable mathematical problem, we first define the distance on any graph
\\
\textbf{Defination 3}: For two vertices $v_1$,$v_2$ in a graph $G(V,E)$, define their distance as the length of the least path connecting the two vertices, written as $d(u_1,u_2)$.
\\
\\
Then we have the following lemma.
\\
\textbf{Lemma 1} : Given a CNOT gate $CNOT_{1,i}$, and denote $d$ to be distance of $1$ and $i$ in a graph $G(V,E)$, then $CNOT_{1,i}$ can be implemented on $G$ with $s$ CNOT gates and depth $s$ without changing any other qubits, where
$$ s=\left\{
\begin{array}{rcl}
4d-4      &      & {d > 1}\\
1     &      & {d = 1}\\
\end{array} \right. $$
\\
$Proof.$ It is trivial for the case of $d = 1$. If $d > 1$, without loss of generality denote $i = d+1$ and $\{ 1,2,...,d+1\}$ to be $d$ points connecting $1$ and $i$, and perform the following procedure.
\\
$(a)$ Perform $CNOT_{j,j+1}$, $j$ from $d$ decrease to $1$. ($d$ depth)
\\
$(b)$ Perform $CNOT_{j,j+1}$, $j$ from $2$ increase to $d$. ($d-1$ depth)
\\
Then the input of vertex $1$ is added to all other vertices.
\\
$(c)$ Perform $CNOT_{j,j+1}$, $j$ from $d-1$ decrease to $1$. ($d-1$ depth)
\\
$(d)$ Perform $CNOT_{j,j+1}$, $j$ from $2$ increase to $d-1$. ($d-2$ depth)
\\
Thus, we implement  $CNOT_{1,i}$ without changing any other bits. \hfill $\Box$
\\
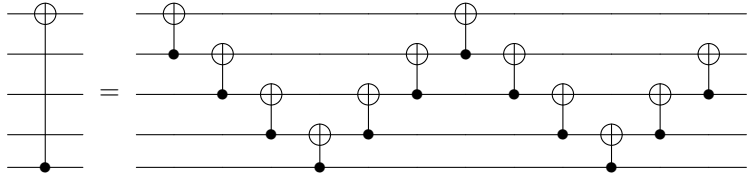
\begin{figure}[H]
$$
\Qcircuit @C=1em @R=0.7em {
&\targ &\qw & \qquad  & \qquad& \targ & \qw & \qw & \qw & \qw & \qw & \targ & \qw & \qw & \qw & \qw & \qw &\qw \\
&\qw&\qw &\qquad  & \qquad& \ctrl{-1} & \targ & \qw & \qw & \qw & \targ & \ctrl{-1} & \targ & \qw & \qw & \qw & \targ &\qw  \\
&\qw &\qw&=  & \qquad& \qw & \ctrl{-1} & \targ & \qw & \targ  & \ctrl{-1} & \qw & \ctrl{-1} & \targ & \qw & \targ & \ctrl{-1} &\qw \\
&\qw &\qw& \qquad  & \qquad& \qw & \qw & \ctrl{-1} & \targ& \ctrl{-1}& \qw & \qw & \qw & \ctrl{-1} & \targ & \ctrl{-1}  & \qw &\qw \\
&\ctrl{-4}&\qw &\qquad & \qquad & \qw& \qw & \qw & \ctrl{-1}& \qw & \qw  & \qw & \qw & \qw & \ctrl{-1}  & \qw & \qw &\qw \\
}
$$
\caption{an example of CNOT circuit between two qubits of distance 4}\label{fig:4}
\end{figure}
Lemma 1 guarantees that for any CNOT circuit, as long as our topological constraint is a connected graph, the quantum circuit can be transformed into an equivalent quantum circuit that can be implemented under the topological constraint.
\\
\\
\textbf{Definition 4}: Given an $n \times n$ matrix $A$ over $GF(2)$ and a representation of $A$ as product of some elementary matrices $E(i+j)$ and a permutation matrix $A'$, denote $S$ as set of these elementary matrices. Define a multigraph $G(V,E,A)$ corresponds to $A$, where $|V|=n$, $E=\{ (v_i,v_j)|E(i+j) \ or \ E(j+i)\in S\}$.
\\
\\
Thus we describe the matrix implementation of a CNOT circuit as a graph $G$, each vertex in the graph corresponds to a qubit, then the problem becomes how to match $G$ to $G'$ such that the cost of implementing all CNOT gates in $G'$ is minimal, where $G'$ is the topological constraint of quantum memories.
\\
\\
\textbf{Problem 3}: Given $G_1(V_1, E_1)$, $G_2(V_2, E_2)$, $|V_1|=|V_2|$, finding a mapping from $G_1$ to $G_2$, $G_1 \rightarrow G_2,v_i \rightarrow x_i$, such that the sum $\sum_{(v_i,v_j)\in E_1} s(x_i,x_j)$ is the minimum, where $s$ is defined in Lemma 1.
\\
\\
Unfortunately, there isn't an algorithm to obtain the theoretical optimal solution for this problem at present. Here we put forward an optimization algorithm to find a solution. Our idea is to transform this problem into a integer programming problem($IP$), then use the solver Gurobi to solve it. We further optimize the results of Gurobi and finally get a better result.
\\
First of all, we define distance functions for different topological constraints. Here we cansider several different architectures, such as 16q-square and IBM 20 in Figure 7.
\begin{figure}[htb]
$$\begin{tikzpicture}
\draw[step=1,color=black, very thick] (0,0) grid (3,3);
\draw[step=1,color=black, very thick] (5,0) grid (9,3);
\draw[color=black, very thick] (5,1)--(7,3);
\draw[color=black, very thick] (6,0)--(9,3);
\draw[color=black, very thick] (9,1)--(8,0);
\draw[color=black, very thick] (9,2)--(8,3);
\draw[color=black, very thick] (9,0)--(6,3);
\draw[color=black, very thick] (5,2)--(7,0);
\fill (0,0) circle(.1);\fill (0,1) circle(.1);\fill (1,0) circle(.1);\fill (1,1) circle(.1);\fill (2,0) circle(.1);\fill (2,1) circle(.1);\fill (3,0) circle(.1);\fill (3,1) circle(.1);\fill (0,2) circle(.1);\fill (0,3) circle(.1);\fill (1,2) circle(.1);\fill (1,3) circle(.1);\fill (2,2) circle(.1);\fill (2,3) circle(.1);\fill (3,3) circle(.1);\fill (3,2) circle(.1);
\node[fill=blue!5, thick ] at (0,0) {$1$};
\node[fill=blue!5, thick ] at (1,0) {$2$};
\node[fill=blue!5, thick ] at (2,0) {$3$};
\node[fill=blue!5, thick ] at (3,0) {$4$};
\node[fill=blue!5, thick ] at (3,1) {$8$};
\node[fill=blue!5, thick ] at (2,1) {$7$};
\node[fill=blue!5, thick ] at (1,1) {$ 6$};
\node[fill=blue!5, thick ] at (0,1) {$ 5$};
\node[fill=blue!5, thick ] at (0,2) {$9$};
\node[fill=blue!5, thick ] at (1,2) {$10$};
\node[fill=blue!5, thick ] at (2,2) {$11$};
\node[fill=blue!5, thick ] at (3,2) {$12$};
\node[fill=blue!5, thick ] at (3,3) {$16$};
\node[fill=blue!5, thick ] at (2,3) {$15$};
\node[fill=blue!5, thick ] at (1,3) {$14$};
\node[fill=blue!5, thick ] at (0,3) {$13$};

\node[fill=blue!5, thick ] at (5,0) {$1$};
\node[fill=blue!5, thick ] at (6,0) {$2$};
\node[fill=blue!5, thick ] at (7,0) {$3$};
\node[fill=blue!5, thick ] at (8,0) {$4$};
\node[fill=blue!5, thick ] at (9,0) {$5$};
\node[fill=blue!5, thick ] at (9,1) {$6$};
\node[fill=blue!5, thick ] at (8,1) {$ 7$};
\node[fill=blue!5, thick ] at (7,1) {$ 8$};
\node[fill=blue!5, thick ] at (6,1) {$9$};
\node[fill=blue!5, thick ] at (5,1) {$10$};
\node[fill=blue!5, thick ] at (5,2) {$11$};
\node[fill=blue!5, thick ] at (6,2) {$12$};
\node[fill=blue!5, thick ] at (7,2) {$13$};
\node[fill=blue!5, thick ] at (8,2) {$14$};
\node[fill=blue!5, thick ] at (9,2) {$15$};
\node[fill=blue!5, thick ] at (9,3) {$16$};
\node[fill=blue!5, thick ] at (8,3) {$ 17$};
\node[fill=blue!5, thick ] at (7,3) {$ 18$};
\node[fill=blue!5, thick ] at (6,3) {$19$};
\node[fill=blue!5, thick ] at (5,3) {$20$};

\end{tikzpicture}$$
\caption{16q-square and IBM 20}\label{fig:5}
\end{figure}
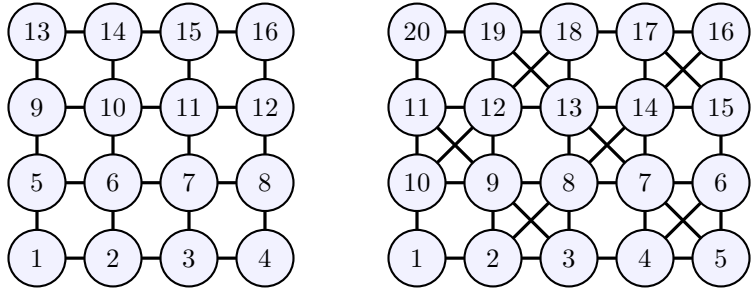
\begin{figure}[htb]
\centering
\includegraphics[scale=0.6]{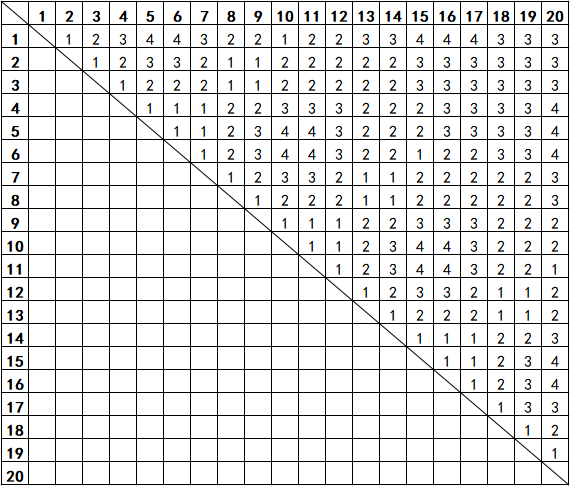}
\caption{Distance function of IBM 20} \label{fig:2}
\end{figure}

For vertice in 16q-square, we can easliy check that
$$ d(x_i,x_j) = |(x_i-1)\ mod\ 4-(x_j-1)\ mod \ 4|+|\lfloor(x_i-1)/4\lfloor-\lfloor(x_j-1)/4\lfloor|$$

But in some other architectures such as IBM 20, we have to define the function as a table in Figure 8.

Now we can write our objective function $\sum_{(v_i,v_j)\in E_1} s(x_i,x_j)$ as a mathematical expression, where $x_i \in [1,n] $ and $x_i \neq x_j$ if $i \neq j$. Minimizing this function is a typical MILP problem. For MILP problems, there are relatively mature solvers such as Gurobi and SAT. Here we use the former. Puting our objective function into Gurobi with , we can get the optimal solution under the current running time based on algorithms in Gurobi with \textbf{Algorithm1}. If the running time was infinite, we would get the optimal solution. But it is impossible for most practical problems. We can only run Gurobi for a certain amount of time and then get a current optimal solution.
\begin{algorithm} 
 [h]
 \caption{Build and solve MILP models}
 \label{alg1}
 \hspace*{0.02in}{\bf Input:}
 Point set $X$\\
 \hspace*{0.02in}{\bf Output:} 
Minimum distance
 
 \begin{algorithmic}[2]

  \Procedure{ConstructMILPModel}{$X$}
  \State Precompute the full linear integer  inequality characterization of distance function;
  \State Initialize an empty Model $\mathcal{M}$;
  \State  $\mathcal{M}.addVar(x_{i,j})$, $i\in[1,n], j\in[0,t]$, where $x_{i,j} $ are binary variables;
  \State  $\mathcal{M}.addVar(s_{i,j}^k)$, $(i,j)\in E_1, k\in[0,t]$, where $s_{i,j}^k $ are binary variables;
  \For{$i,j\in [1,n]$, $i\neq j$}
  \State $x_i=\sum_{k=0}^{t} 2^k{x_{i,k}}$;\Comment{Binary to integer representation}
  \State $x_j=\sum_{k=0}^{t} 2^k{x_{j,k}}$;
  \State  $\mathcal{M}.addConstr (abs(x_i-x_j)\geq 1)$;\Comment{distinct}
  \State  Adding constrains of $x_{i}, x_{j}$ and $s_{i,j}^k $ to $\mathcal{M}$ according to the full linear integer  inequality characterization of distance function;
  \EndFor
  \State $\mathcal{M}.setObjective \gets Minimize \sum_{i,j} \sum _k2^k\cdot s_{i,j}^k$; 
  \State $\mathcal{M}.Optimize()$;
  \State $D\gets $ the best solution of $\mathcal{M}$;
  
  \Return{$D$};
  \EndProcedure
 \label{code:recentEnd} 
 \end{algorithmic}
\end{algorithm}

As for the results of Gurobi, it is natural to think whether further optimization can be done. Our idea  is permutation of the points. We exchanged the grid points on the rectangular lattice to calculate the objective function. If it is reduced, the current result is replaced with the result after exchange. The number of points exchanged can be changed based on the different scale of the problem. Specific details are shown in \textbf{Algorithm 2}.
\begin{algorithm}
  \caption{Permutation of points}  
\hspace*{0.02in}{\bf Input:}
 $\{x_i\}, i \in [1,n], x_i \in [1,n], x_i \neq x_j\  if \ i \neq j$, $Objective function \ \ S$
  \begin{algorithmic}[2]  
	\Procedure{Permutation}{$\{x_i\}$}
    \State Calculate $S=S_{initial}=\sum_{(v_i,v_j)\in E} s(x_i,x_j)$
    \For{each subset of order k $D_k=\{i_1,i_2,...,i_k\} \subseteq [1,n]$}  
    \For{each $\sigma \in S_k$,where $S_k$ is a permutation group on $D_k$}
		\State Calculate $S'=\sum_{(v_i,v_j)\in E} s(\sigma(x_i),\sigma(x_j))$,where $\sigma(x_i)=x_i$ if $x_i \notin D_k$
		\If{$S'<S_{initial}$}
		\State $x_i = \sigma(x_i)  \quad S=S'$ 
		\EndIf
    \EndFor  
     \EndFor
\While{$S<S_{initial}$}
\State Repeat $Step \ 2 \sim 10$
\EndWhile
  	\State \Return $\{x_i\}$
\EndProcedure
    \label{code:recentEnd}  
  \end{algorithmic}  
\end{algorithm}

Here we take a $5\times 5$ matrix in \cite{2019arXiv191014478W} as an example. The matrix is same as $A$ in section 3.1.1 and the topological constraint graph $G$ is shown in Figure 9. By decomposition of $A$, $G(V,E,A)$ is a graph as in Figure 9. The number of points is so small that we can get the optimal solution only by algorithm 2. Then our objective function $\sum_{(v_i,v_j)\in G(V,E,A)} s(x_i,x_j)$ goes to a minimum of 16 in Figure 10.
\begin{figure}[H]
$$\begin{tikzpicture}
\draw[color=black, very thick] (1,0)--(1,2);
\draw[color=black, very thick] (0,2)--(2,2);
\draw[color=black, very thick] (4,1)--(5,2);
\draw[color=black, very thick] (5,2)--(6,2);
\draw[color=black, very thick] (5,2)--(5,0);
\draw[color=black, very thick] (5,0)--(6,2);
\draw[color=black, very thick] (5,0)--(4,1);
\draw[color=black, very thick] (5,0)--(6,0);
\draw[color=black, very thick] (4,1)--(6,0);
\node[fill=blue!5, thick ] at (1,0) {$1$};
\node[fill=blue!5, thick ] at (1,1) {$2$};
\node[fill=blue!5, thick ] at (1,2) {$3$};
\node[fill=blue!5, thick ] at (0,2) {$4$};
\node[fill=blue!5, thick ] at (2,2) {$5$};
\node[fill=blue!5, thick ] at (4,1) {$v_1$};
\node[fill=blue!5, thick ] at (5,0) {$v_2$};
\node[fill=blue!5, thick ] at (5,2) {$v_3$};
\node[fill=blue!5, thick ] at (6,2) {$v_4$};
\node[fill=blue!5, thick ] at (6,0) {$v_5$};
\end{tikzpicture}$$
\caption{The topological constraint $G$ and representation of $A$, $G(V,E,A)$}\label{fig:5}
\end{figure}
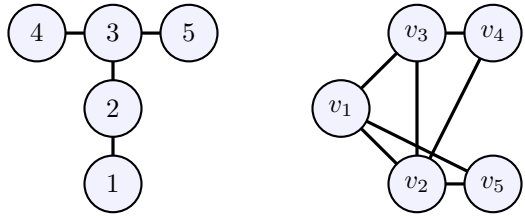
\begin{figure}[H]
$$\begin{tikzpicture}
\draw[color=black, very thick] (1,0)--(1,2);
\draw[color=black, very thick] (0,2)--(2,2);
\node[fill=blue!5, thick ] at (1,1) {$v_1$};
\node[fill=blue!5, thick ] at (1,2) {$v_2$};
\node[fill=blue!5, thick ] at (2,2) {$v_3$};
\node[fill=blue!5, thick ] at (0,2) {$v_4$};
\node[fill=blue!5, thick ] at (1,0) {$v_5$};
\end{tikzpicture}$$
\caption{The map between $G(V,E,A)$ and $G$}\label{fig:5}
\end{figure}
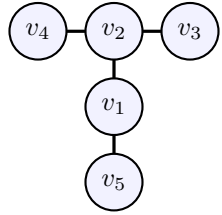

Then we can get a $CNOT$ circuit of A under topological constraint of $G$ with $size=16$ as in Figure 11. It is less 4 than the example of method ROWCOL in \cite{2019arXiv191014478W}.
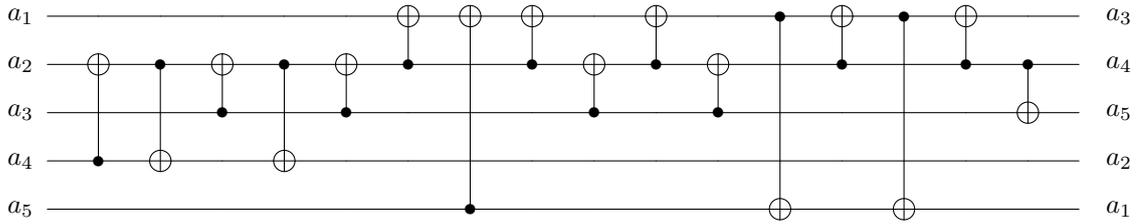
\begin{figure}[H] 
\centering
$$\Qcircuit @C=1.5em @R=1em { 
&a_1 \qquad & \qw       & \qw       	&\qw  	&\qw  		 &\qw				&\targ      & \targ     & \targ       &\qw 		& \targ      &\qw       & \ctrl{4} & \targ & \ctrl{4}  & \targ  & \qw      & \qw &a_3 \\
&a_2 \qquad & \targ     & \ctrl{2}   &\targ     & \ctrl{2} 	 &\targ			& \ctrl{-1} & \qw       & \ctrl{-1}   &\targ  		& \ctrl{-1}  &\targ     & \qw       & \ctrl{-1}  & \qw    & \ctrl{-1}   & \ctrl{1} & \qw &a_4\\
&a_3 \qquad & \qw       & \qw     	&\ctrl{-1}	&\qw 			 &\ctrl{-1}		& \qw       & \qw       & \qw		 &\ctrl{-1} 	&\qw         &\ctrl{-1} & \qw      &\qw &\qw     & \qw  & \targ    & \qw &a_5\\
&a_4 \qquad & \ctrl{-2} & \targ 	&\qw 		&\targ 		 &\qw				& \qw       & \qw       & \qw         & \qw      	& \qw        & \qw      &\qw   & \qw & \qw     	  &\qw  &\qw       &\qw &a_2\\
&a_5 \qquad & \qw       & \qw        & \qw 	& \qw 		 & \qw			& \qw       & \ctrl{-4} & \qw         &\qw  		&\qw	        &\qw        & \targ    &\qw	  & \targ   & \qw     & \qw & \qw &a_1\\
}$$

\caption{the CNOT circuit of $A$ with topological constraints}\label{fig:1}
\end{figure}

\section{Experiments}
\subsection{Comparison with other methods}
We compare our algorithm with the most commonly used Steiner Tree algorithm. For the sake of justice, we use the same matrixes with \cite{kissinger2019cnot}, the number of each kind of CNOT circuit is 20, and we use a test set of 380 random circuits in total. The result is listed in Table 1.
\begin{table}[h]
\centering
\begin{tabular}{|c|c|c|c|c|c|c|}
\hline
Architecture & \#  & QuilC  & t|ket\textgreater{} & Steiner & our method & \textless{}Steiner \\
\hline
9q-square    & 3  & 3.8    & 3.6                  & 3       &     2.95   &           1.6\%         \\
9q-square    & 5   & 10.82  & 6.4                 & 5.2     &    4.6     &           11.5\%         \\
9q-square    & 10  & 20.08  & 16.95               & 11.6    &    11.7    &              -0.9\%      \\
9q-square    & 20  & 46.24  & 40.75               & 25.85   &    23.8    &                7.9\%    \\
9q-square    & 30  & 72.89  & 66.15               & 35.55   &    31.3    &               12.0\%     \\
16q-square   & 4   & 6.14   & 5.8                 & 4.44    &      4     &                10.0\%   \\
16q-square   & 8   & 19.68  & 12.95               & 12.41   &     7.65   &               38.4\%     \\
16q-square   & 16  & 48.13  & 36.2                & 33.08   &       25   &              24.4\%      \\
16q-square   & 32  & 106.75 & 94.45               & 82.95   &    68.5    &             17.4\%       \\
16q-square   & 64  & 225.69 & 203.75              & 147.38  &      138.15&               6.3\%     \\
16q-square   & 128 & 457.35 & 436.25              & 168.12  &      150.25&               10.6\%     \\
16q-square   & 256 & 925.85 & 922.65              & 169.28  &      153.65&                9.2\%    \\
ibm-q20-tokyo & 4   & 5.5    & 6.05   		  & 4     	 &  		4   &  0 \\
ibm-q20-tokyo & 8   & 17.3   & 12                  & 7.69   &        7.8  &   -1.4\%\\
ibm-q20-tokyo & 16  & 43.83  & 29.05             & 20.44 	  &       14.85  &  27.3\%\\
ibm-q20-tokyo & 32  & 93.58  & 78.15            & 66.93  	  &      49.35      &  26.3\%\\
ibm-q20-tokyo & 64  & 215.9  & 181.25           & 165.6      &      124.2     & 25.0\% \\
ibm-q20-tokyo & 128 & 432.65 & 391.85           & 237.64     &       217.95    &  8.3\%\\
ibm-q20-tokyo & 256 & 860.74 & 789.3             & 245.84    &       219.5     & 10.7\%\\ \hline
\end{tabular}
\caption{The second colunmn shows the number of CNOT gates of random circuits. The remaining columns show the average
2-qubit gate count after mapping 20 random circuits. }\label{fig:1}
\end{table}

\subsection{Application in AES's Mixcolumns}
The quantum circuit implementation of AES is an important problem\cite{bonnetain2019quantum} \cite{grassl2016applying}, but most of the researchers neglects the structure of quantum computer. They acquiesced that the cost of AES's Mixcolumns is 92 CNOT gates without constrain which seems a little unreasonable on real quantum computers. In this section we apply our method in the MC layer of AES which is written as a $32 \times 32$ matrix. We use a $4 \times 8$ lattice as topologically-constrained quantum memories, and get a result of 159 rather than 92 CNOT gates of the cost of AES's Mixcolumns. Here is our result.
\begin{figure}[h]
$$\begin{tikzpicture}
\draw[step=1,color=gray!40] (0,0) grid (7,3);
\node[fill=blue!5, thick ] at (0,0) {$v_3$};
\node[fill=blue!5, thick ] at (1,0) {$v_{11}$};
\node[fill=blue!5, thick ] at (2,0) {$v_{12}$};
\node[fill=blue!5, thick ] at (3,0) {$v_{28}$};
\node[fill=blue!5, thick ] at (4,0) {$v_{21}$};
\node[fill=blue!5, thick ] at (5,0) {$v_5$};
\node[fill=blue!5, thick ] at (6,0) {$v_{30}$};
\node[fill=blue!5, thick ] at (7,0) {$v_6$};
\node[fill=blue!5, thick ] at (0,1) {$v_{27}$};
\node[fill=blue!5, thick ] at (1,1) {$v_{10}$};
\node[fill=blue!5, thick ] at (2,1) {$v_4$};
\node[fill=blue!5, thick ] at (3,1) {$v_{20}$};
\node[fill=blue!5, thick ] at (4,1) {$v_{13}$};
\node[fill=blue!5, thick ] at (5,1) {$v_{29}$};
\node[fill=blue!5, thick ] at (6,1) {$v_{22}$};
\node[fill=blue!5, thick ] at (7,1) {$v_{14}$};
\node[fill=blue!5, thick ] at (0,2) {$v_{19}$};
\node[fill=blue!5, thick ] at (1,2) {$v_{18}$};
\node[fill=blue!5, thick ] at (2,2) {$v_{1}$};
\node[fill=blue!5, thick ] at (3,2) {$v_{17}$};
\node[fill=blue!5, thick ] at (4,2) {$v_{24}$};
\node[fill=blue!5, thick ] at (5,2) {$v_{32}$};
\node[fill=blue!5, thick ] at (6,2) {$v_{23}$};
\node[fill=blue!5, thick ] at (7,2) {$v_{7}$};
\node[fill=blue!5, thick ] at (0,3) {$v_2$};
\node[fill=blue!5, thick ] at (1,3) {$v_{26}$};
\node[fill=blue!5, thick ] at (2,3) {$v_9$};
\node[fill=blue!5, thick ] at (3,3) {$v_{25}$};
\node[fill=blue!5, thick ] at (4,3) {$v_{16}$};
\node[fill=blue!5, thick ] at (5,3) {$v_8$};
\node[fill=blue!5, thick ] at (6,3) {$v_{15}$};
\node[fill=blue!5, thick ] at (7,3) {$v_{31}$};
\end{tikzpicture}$$
\caption{The routing of 32 qubits in AES's Mixcolumns, Size=159}\label{fig:3}
\end{figure}
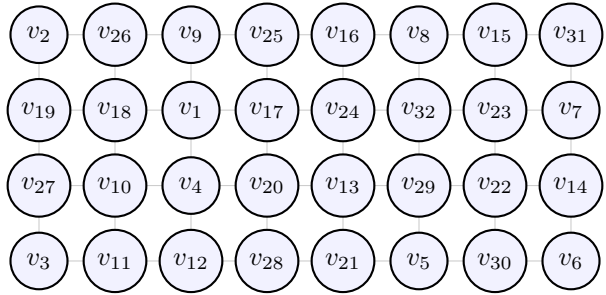
\\
\\

\bibliographystyle{alpha}
\bibliography{abbrev3,crypto,bilbio}

\end{document}